\documentclass[twocolumn,prl,superscriptaddress,float,aps]{revtex4-2}
\usepackage{graphicx,amsfonts,amssymb,amsmath,hyperref,hypcap,enumerate}
\usepackage{color}
\usepackage{bm}
\usepackage{multirow}
\usepackage{makecell}
\usepackage{color}
\usepackage{orcidlink}
\usepackage{hyperref}
\usepackage[title]{appendix}
\hypersetup{
colorlinks = true,
linkcolor = [rgb]{0.70,0.13,0.13},
citecolor = [rgb]{0.13,0.55,0.13},
urlcolor = [rgb]{0.25, 0.41, 0.88}}
\usepackage{makecell}

\newcommand{\eqs}[1]{\begin{equation}\begin{split}#1\end{split}\end{equation}}
\newcommand{\eqnref}[1]{Eq.\,\eqref{#1}}
\newcommand{\figref}[1]{Fig.\,\ref{#1}}

\begin{document}
\title{Deconfined Quantum Critical Point in  Quantum Hall Bilayers}

\author{Guangyu Yu\orcidlink{0009-0000-7405-1853}}
\affiliation{Kavli Institute for Theoretical Sciences, University of Chinese Academy of Sciences, Beijing 100190, China}

\author{Tao Xiang\orcidlink{0000-0001-5998-7338}}
\email{txiang@iphy.ac.cn}
\affiliation{Institute of Physics, Chinese Academy of Sciences, Beijing 100190, China}

\author{Zheng Zhu\orcidlink{0000-0001-7510-9949}}
\email{zhuzheng@ucas.ac.cn}
\affiliation{Kavli Institute for Theoretical Sciences, University of Chinese Academy of Sciences, Beijing 100190, China}

\begin{abstract} 

Deconfined quantum critical points (DQCPs) represent an unconventional class of quantum criticality beyond the Landau-Ginzburg-Wilson-Fisher paradigm. Nevertheless, both their theoretical identification and experimental realization remain challenging.
Here we report compelling evidence of a DQCP in quantum Hall bilayers with half-filled $n=2$ Landau levels in each layer, based on large-scale variational uniform matrix product state (VUMPS) simulations and exact diagonalization (ED).
By systematically analyzing the ground-state fidelity, low-lying energy spectra, exciton superfluid and stripe order parameters, and ground-state energy derivatives, we identify a direct and continuous quantum phase transition 
between two distinct symmetry-breaking phases by tuning the layer separation: an exciton superfluid phase with spontaneous $U(1)$ symmetry breaking at small separation,  and a unidirectional charge density wave with broken translational symmetry at large separation. 
Our results highlight quantum Hall bilayers as an ideal platform for realizing and experimentally probing DQCPs under precisely tunable interactions.

\end{abstract}

\maketitle

\emph{Introduction.---} Quantum phase transition is one of the central focuses in condensed matter physics.  Although the Landau-Ginzburg-Wilson-Fisher (LGWF) theory successfully describes numerous quantum phase transitions governed by single-order-parameter fluctuations, there are certain transitions beyond the LGWF paradigm. Understanding these transitions poses a challenging issue of fundamental importance. A typical category is the deconfined quantum critical point (DQCP) \cite{doi:10.1143/JPSJS.74S.1,doi:10.1126/science.1091806,senthil_2023_deconfined}, which represents a continuous quantum phase transition between two distinct symmetry-breaking phases that do not include each other as a subgroup. These phases are characterized by distinct local Landau order parameters, and thus, direct continuous transitions are 
prohibited within the framework of LGWF theory.

The DQCP is initially proposed as a transition between the Néel and valence bond solid (VBS) ordered phases on the square lattice \cite{doi:10.1126/science.1091806,PhysRevB.70.144407}, where the Néel phase and the VBS phase break the $SU(2)$ spin symmetry and translation symmetry, respectively. Subsequently, substantial theoretical efforts have been devoted to identifying DQCP in various lattice models and examining their critical behaviors \cite{senthil_2023_deconfined,PhysRevLett.98.227202,PhysRevB.80.180414,PhysRevLett.100.017203,PhysRevB.82.155139,PhysRevLett.104.177201,PhysRevB.88.220408,Jiang_2008,PhysRevLett.110.185701,PhysRevX.5.041048,PhysRevLett.115.267203,motrunich2008comparative,PhysRevLett.101.050405,PhysRevB.88.195140,PhysRevLett.101.167205,PhysRevB.80.045112,PhysRevB.82.014429,PhysRevB.92.184413,PhysRevB.106.155131,PhysRevB.99.075103,PhysRevB.72.134502,PhysRevX.9.041037,PhysRevB.97.195115,PhysRevX.7.031051,PhysRevB.70.220403,PhysRevLett.128.087201,10.21468/SciPostPhys.15.5.215,PhysRevB.102.155124,PhysRevB.109.085143,PhysRevB.99.075103,PhysRevB.100.125137}. 
For instance, the superfluid-Mott transition in two-dimensional (2D) lattice boson systems \cite{PhysRevB.72.134502}, the N\'eel-VBS transition in 2D lattice spin models \cite{PhysRevLett.98.227202}, 3D lattice loop models \cite{PhysRevX.5.041048} and 1D spin lattice models \cite{PhysRevB.99.075103,PhysRevB.100.125137}.  
Although the numerical results provide compelling evidence for DQCP, the experimental probe of DQCP based on the proposed lattice models remains elusive \cite{doi:10.1126/science.adc9487}. This challenge is partially due to the complexity of accurately realizing lattice spin models in real materials and the difficulty of accessing sufficiently low temperatures in cold-atom optical lattices. 

With their widely and precisely tunable interactions, quantum Hall systems provide a promising platform for realizing and investigating  DQCP physics. Experimentally, bilayer quantum Hall systems can be realized in single-wide quantum wells \cite{PhysRevLett.68.1379}, double quantum wells \cite{PhysRevLett.68.1383}, and various graphene-based materials \cite{li2017excitonic,liu_watanabe_taniguchi_halperin_kim_2017,liu_hao_watanabe_taniguchi_halperin_kim_2019,li_shi_zeng_watanabe_taniguchi_hone_dean_2019,liu_li_watanabe_taniguchi_hone_halperin_kim_dean_2022,PhysRevLett.129.187701,chen2024tunable}.
In contrast to lattice spin models that require fine-tuned exchange interactions, quantum Hall bilayers \cite{doi:https://doi.org/10.1002/9783527617258.ch5} enable continuous parameter control via layer separation — a knob inaccessible in conventional solid-state systems. This continuously variable layer distance control positions quantum Hall bilayers as a unique platform to realize exceptionally clean quantum phase transitions with minimal parameter engineering \cite{PhysRevX.13.031023,PhysRevLett.131.256502,PhysRevLett.124.097604,PhysRevLett.124.097604,PhysRevB.94.245147,PhysRevLett.87.056802,PhysRevB.85.195113,zibrov_kometter_zhou_spanton_taniguchi_watanabe_zaletel_young_2017,papic:tel-00624760,PhysRevLett.121.026603,PhysRevB.98.045113,hou2025excitoncondensationcompositefermions,PhysRevB.101.085412,PhysRevB.91.205139,PhysRevB.98.155104,PhysRevX.7.041068,han2025anyonsuperfluidityexcitonsquantum}.

\begin{figure}[tpb]
\includegraphics[width=0.9\linewidth]{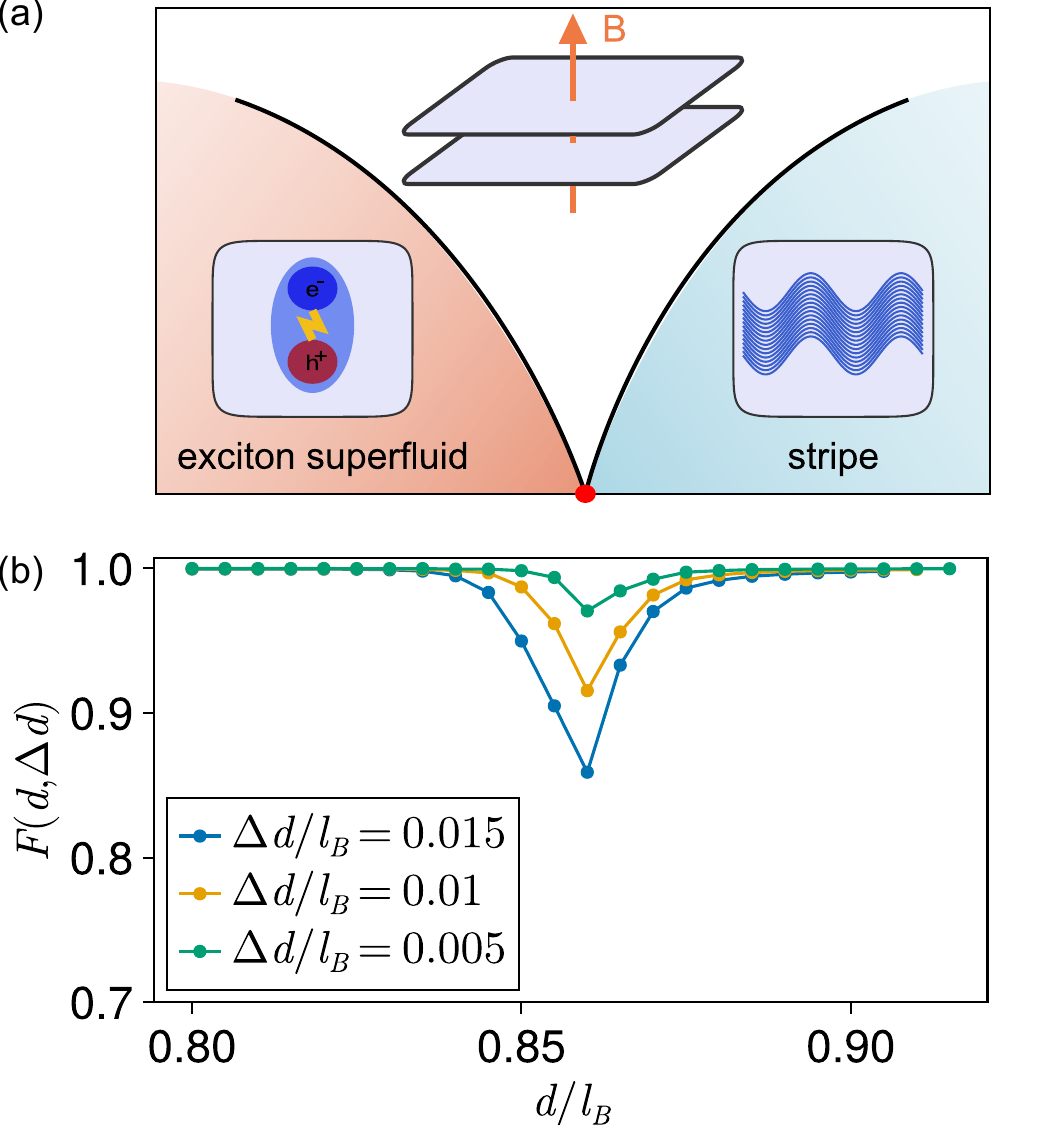}
\caption{(a) Quantum phase transition in quantum Hall bilayers with half-filled $n=2$ Landau levels in each layer.
 The horizontal axis ($d/l_B$) controls interlayer coupling strength. Left (orange): exciton condensate phase with spontaneous $U(1)$ symmetry breaking.  Right (cyan): unidirectional CDW phase with broken translational symmetry.
The critical point realizes a DQCP separating these two symmetry-breaking phases.  (b) VUMPS simulations of ground-state fidelity near the critical layer distance $d_c\approx0.86l_B$. The dip in fidelity weakens as the step size $\Delta d$ decreases, consistent with ED results~\cite{supplement} and demonstrating a continuous quantum phase 
 transition. }
\label{fig:fig0}
\end{figure}

Here we investigate quantum Hall bilayers with a half-filled third Landau level (n=2 LL) in each layer and uncover
a direct and continuous quantum phase transition between the exciton superfluid phase and charge stripe phase upon tuning the layer distances $d/l_B$, where $d$ is the layer distance and $l_B$ is the magnetic length. Using variational uniform matrix product state (VUMPS) method~\cite{PhysRevB.97.045145,10.21468/SciPostPhysLectNotes.7} and exact diagonalization (ED)~\cite{PhysRevLett.55.2095}, we first characterize the exciton superfluid phase at small $d/l_B$ and the unidirectional CDW phase at large $d/l_B$ through numerical diagnostics including the pseudospin excitation gap, energy spectra structure, pseudospin correlations, and the static charge structure factor.
At small $d/l_B$, an electron in one layer binds to a hole in the other layer, forming a bosonic exciton. 
These excitons condense into a superfluid ground state with spontaneous $U(1)$ symmetry breaking
~\cite{eisenstein_2014_exciton,PhysRevLett.69.1811,PhysRevB.51.5138,PhysRevB.54.11644,PhysRevLett.86.1825,PhysRevLett.86.1829,PhysRevLett.119.177601,PhysRevB.99.201108,PhysRevB.54.11644,PhysRevLett.91.116802,eisenstein_macdonald_2004}. In the decoupled limit, the ground state of each layer is believed to be the unidirectional charge density wave (CDW), also known as the stripe phase \cite{PhysRevLett.82.394,PhysRevLett.83.1219}, which spontaneously breaks the translation symmetry. 
Crucially, these two phases break distinct symmetries ($U(1)$ vs. translation) that cannot be unified within a single LGWF framework.
By tuning the layer distances $d/l_B$, we identify a direct quantum phase transition between these two conventional symmetry-breaking phases [see \figref{fig:fig0}(a) for a schematic diagram]. Notably, based on the smooth evolution of low-lying energy spectra, the continuous vanishing of the exciton superfluid order parameter, the absence of singularities in ground-state energy derivatives, and the ground-state fidelity analysis, we demonstrate that this direct quantum phase transition is continuous. These observations are consistent with the phenomenology of DQCP. 
Given that layer separation $d/l_B$ can be tuned precisely by simultaneously adjusting the magnetic field and electron density while maintaining a constant filling factor, this continuous transition is experimentally realizable.

\begin{figure}
\includegraphics[width=1\linewidth]{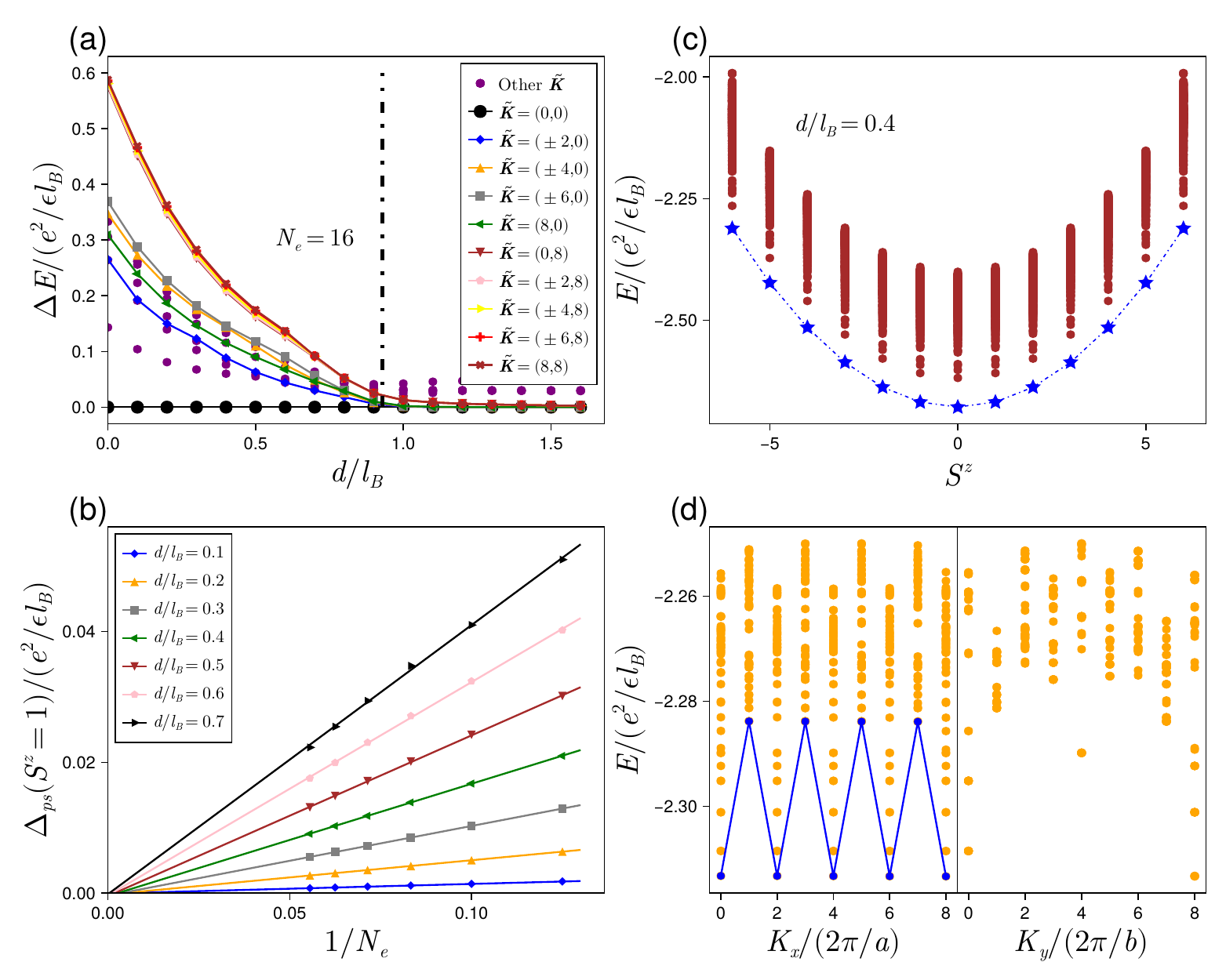}
\caption{(a) Low-lying energy spectra flow with layer distance $d/l_B$. The smooth evolution without level crossing suggests a continuous quantum phase transition at the critical point $d_c/l_B\approx 0.93$ (dash-dotted line) by ED on the $N_e=16$ system.
Here, the ground state consistently resides in the $\mathbf{K}_0$ sector for a system with 16 electrons, where $\mathbf{K}_0=(8,8)$ in the unit of $(2\pi/a,2\pi/b)$ and $\tilde{\mathbf{K}}=\mathbf{K}-\mathbf{K}_0$. 
(b) Finite-size scaling of the pseudospin gap $\Delta_{\mathrm{ps}}$ if we view two layers as pseudospins. $\Delta_{\mathrm{ps}}$ is calculated up to systems with $N_e=18$ electrons and extrapolates to zero at $d<d_c$, consistent with exciton superfluid, or equivalently, the pseudospin ferromagnetic long-range order. (c) Low-lying energy spectra in different pseudospin $S^z$ sectors at $d/l_B=0.4$, indicating an easy-plane ferromagnet at $d<d_c$. An energy shift $d\cdot S^2_z /N_\phi$ due to layer charge imbalance is subtracted. (d) Energy spectra versus the momentum $K_x$ and $K_y$ for systems with $N_e=16$ electrons at layer distance $d/l_B=1.3$. Distinct structure implies the presence of a unidirectional CDW phase.  
 }
\label{fig:fig1}
\end{figure}

\emph{Model and Method.---} We consider the quantum Hall bilayer with half-filled n=2 LL in each identical layer. We assume full spin polarization and negligible Landau level mixing under strong magnetic fields. Due to quenched kinetic energy by the strong magnetic field, the system is governed by projected Coulomb interactions:
\eqs{\label{eq:H}H=\sum_{\substack{i<j\\ \alpha,\beta}}\sum_{\mathbf{q}}\frac{V_{\alpha\beta}(\mathbf{q})}{A}e^{-\frac{q^2l_B^2}{2}}L^2_n(\frac{q^2 l_B^2}{2})e^{i\mathbf{q}\cdot(\mathbf{R}_{\alpha,i}-\mathbf{R}_{\beta,j})}}
where $\alpha(\beta)=\uparrow,\downarrow$ denotes the layer index or equivalently the pseudospin index, $q=|\mathbf{q}|=\sqrt{q_x^2+q_y^2}$, $ V_{\uparrow\uparrow}(\mathbf{q})=V_{\downarrow\downarrow}(\mathbf{q})=2\pi e^2/(\epsilon q)$ and $V_{\uparrow\downarrow}(\mathbf{q})=V_{\downarrow\uparrow}(\mathbf{q})=2\pi e^2/(\epsilon q)\cdot e^{-qd/l_B}$ are the Fourier transformations of the intralayer and interlayer Coulomb interactions, respectively. $d$ represents the layer distance, $\epsilon$ is the dielectric constant and $A$ is the area of the system. $L_n(x)$ are the Laguerre polynomials, and $\mathbf{R}_{\alpha,i}$ is the guiding center coordinate of the $i$-th electron in layer $\alpha$. 
 
We adopt the Landau gauge, with geometries tailored to each numerical approach: ED is performed on a torus defined by lattice vectors $\mathbf{a}$ and $\mathbf{b}$ with aspect ratio $b/a=0.64$~\cite{PhysRevLett.83.1219}, while VUMPS calculations utilize an infinite cylinder geometry with $L_y=12l_B$. The flux number (or orbital number) in each layer $N_\phi$ satisfies $A=|\mathbf{a}\times \mathbf{b}|=2\pi l_B^2 N_\phi$ for toroidal ED systems, where $l_B=\sqrt{{\hbar c}/{eB}}$ denotes the magnetic length. Throughout this work, we employ $l_B$ as the unit of length and $e^2/{\epsilon l_B}$ as the energy unit. To eliminate the potential finite-size effect in ED, we employ the VUMPS algorithm to examine the robustness of conclusions in the thermodynamic limit. Compared to conventional infinite density matrix renormalization group (iDMRG) approaches, VUMPS demonstrates superior convergence in capturing ground-state properties \cite{PhysRevB.97.045145}. To ensure proper selection of the symmetry sector and numerical stability, we implement the controlled bond expansion (CBE) algorithm \cite{PhysRevLett.130.246402} during optimization. For computational feasibility in infinite-size simulations, we adopt a modified Coulomb interaction with Gaussian regularization \cite{PhysRevB.91.045115,PhysRevLett.115.126805}: $V_{\rm{intra}}(r)= \frac{e^{-r^2/\xi^2}}{r}$, $V_{\rm{inter}}(r)= \frac{e^{-r^2/\xi^2}}{\sqrt{r^2+d^2}}$, where the cutoff length $\xi = 6l_B$ ensures smooth decay while preserving universal critical behavior. This regularization only affects non-universal quantities like the precise critical point location rather than the critical behavior. We perform simulations up to bond dimension $\chi=4000$ in a 12-site unit cell configuration, yielding well-converged results (details see Supplementary Material~\cite{supplement}).

\begin{figure}
\includegraphics[width=1\linewidth]{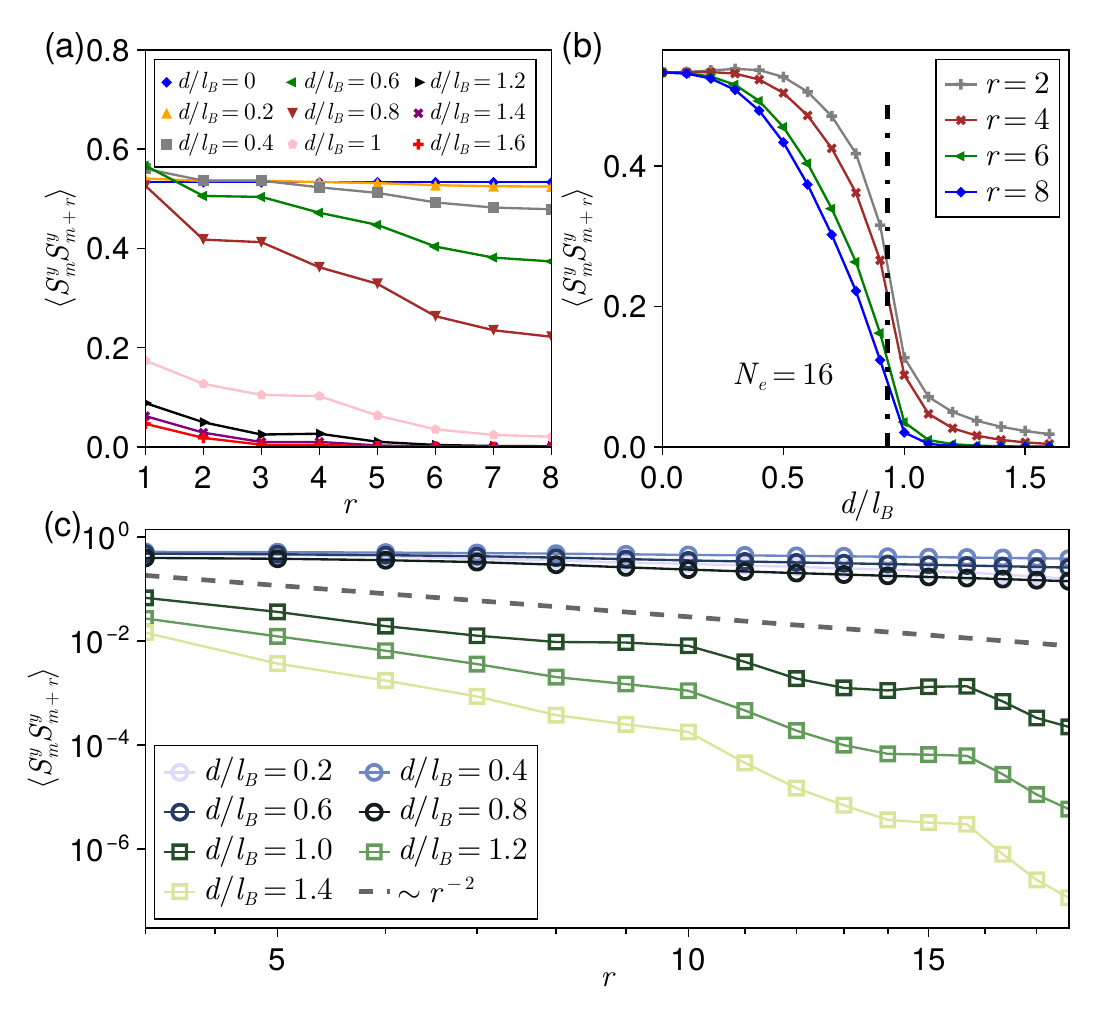}
\caption{(a) Pseudospin correlation as a function of orbital separation $r$ at various layer distances for systems with $N_e=16$ electrons. The correlation functions exhibit distinct behaviors across the quantum critical point. (b) Pseudospin correlation $\langle S^y_m S^y_{m+r}\rangle$ versus layer distance at several fixed orbital separations $r$. The correlation smoothly vanishes beyond the critical layer separation, indicated by the dash-dotted line. (c) VUMPS-calculated pseudospin correlation on infinite cylinders.
The correlations decay slowly or remain finite at $d < d_c$, while they decay faster than $r^{-2}$ (gray dashed line) at $d > d_c$, indicating the absence of interlayer coherence. 
}
\label{fig:fig2}
\end{figure}

\emph{Exciton superfluid phase at $d<d_c$.---} We begin by examining the evolution of low-lying energy spectra as a function of layer distance $d/l_B$ by ED.  As shown in \figref{fig:fig1}(a), the spectra evolve smoothly without level crossing as $d/l_B$ increases, signifying a direct and continuous quantum phase transition at the critical distance $d_c$.
Subsequently, we demonstrate that this transition connects two phases with distinctly broken symmetry.
We first reveal that the phase at $d<d_c$ is the exciton superfluid phase, as evidenced by a vanishing pseudospin excitation gap, distinctive structure in the low-lying energy spectra, and long-range pseudospin correlations.

The pseudospin gap is defined as $\Delta_{ps}(d)\equiv E_0(N_\uparrow,N_\downarrow,d)-E_0(N_{\phi}/2,N_{\phi}/2,d)+d\cdot S_z^2/N_{\phi}$, where $N_\uparrow=N_{\phi}/2+S_z$ and $N_\downarrow=N_\phi/2-S_z$ denote electron numbers in two layers for pseudospin $S_z$. 
The additional term $d\cdot S_z^2/N_{\phi}$ is due to the positive electric background and charge imbalance \cite{PhysRevLett.65.775}.
When the exciton condenses into a superfluid phase, the pseudospin gap vanishes, suggesting that the electrons can tunnel between the two layers without a driving bias voltage. As shown in \figref{fig:fig1}(b), the pseudospin gap extrapolates to zero in the thermodynamic limit at $d<d_c$. This signature explains the resonant peak observed in the interlayer tunneling conductance at zero bias in experiments \cite{eisenstein_2014_exciton}, serving as a robust indicator of the interlayer exciton superfluidity. 
Furthermore, \figref{fig:fig1}(c) reveals a nondegenerate ground state within the $S_z
=0$ sector at small but non-zero layer separations. The low-energy excitations are pseudospin excitations among different $S_z$ sectors and can be accurately fitted to $\Delta E=E(S_z)-E(S_z=0)=\alpha S_z^2$, suggesting that the ground state is an easy-plane ferromagnet instead of an Ising ferromagnet, consistent with exciton superfluid.
 
Next, we probe exciton coherence via the pseudospin correlation function $\langle S^y_m S^y_{m+r}\rangle$, where the pseudospin $S^y_m\equiv i(c_{\uparrow m}^\dagger c_{\downarrow m}-c_{\downarrow m}^\dagger c_{\uparrow m})$ represents the exciton superfluid order parameter, (and is often referred to as the current in experimental contexts), and $m=1,...,N_\phi$ denotes orbital indices~\cite{PhysRevX.13.031023,PhysRevB.99.201108}. 
As shown in \figref{fig:fig2}(a), the pseudospin correlations decay slowly and saturate to a finite value at $d<d_c$, demonstrating the existence of interlayer coherence. In contrast, it has a much smaller amplitude and rapidly vanishes at larger distances. We further employ VUMPS to verify the robustness of ED results in the thermodynamic limit. As shown in \figref{fig:fig2}(c), the pseudospin correlations display contrasting decay behaviors across the critical layer distance $d_c$. For $d < d_c$, the correlations either remain finite or decay much slower than $r^{-2}$, indicating a divergent susceptibility in 2D as the temperature $T \to 0$. However, for $d > d_c$, the correlations decay much faster than $r^{-2}$, suggesting the absence of interlayer coherence. These numerical observations suggest that the exciton superfluid remains robust in the thermodynamic limit.

To capture the evolution of such properties with increasing layer distance, we focus on the pseudospin correlations $\langle S^y_m S^y_{m+r}\rangle$ at several fixed orbital distances $r$. As shown in \figref{fig:fig2}(b), the pseudospin correlations remain finite at a small layer distance $d<d_c$, and smoothly decay to zero after crossing the critical layer distance, suggesting a continuous quantum phase transition.

\begin{figure}[]
\includegraphics[width=1\linewidth]{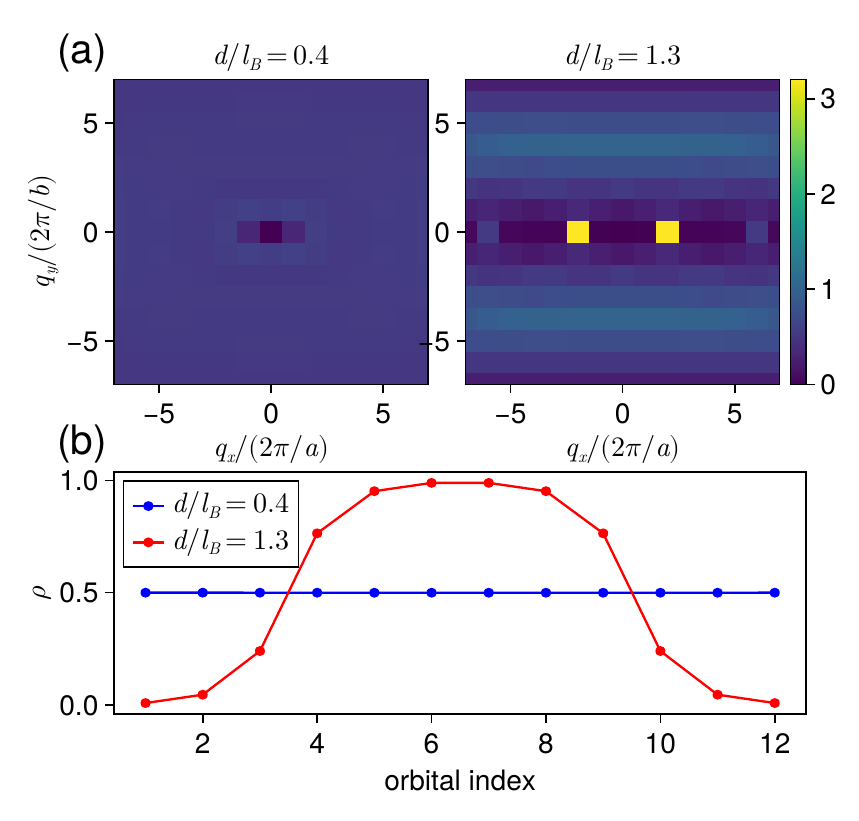}
\caption{(a) Charge guiding-center structure factor $S(\mathbf{q})$ at representative  layer separations $d/l_B$ for systems with $N_e = 16$ electrons. $S(\mathbf{q})$ remains uniform without significant peaks at $d<d_c$ (left) and exhibits two sharp peaks at $d>d_c$ (right), suggesting the unidirectional charge ordering. The prominent peak is located at $\mathbf{q}^* = (2\cdot\frac{2\pi}{a},0)$ at $d>d_c$. (b) VUMPS-calculated charge density profile of one layer in one unit cell along the cylinder. }
\label{fig:fig3}
\end{figure}

\emph{Charge stripe phase at $d>d_c$.---}
We first examine the structure of the low-energy spectra, as illustrated in \figref{fig:fig1}(d) for a layer distance of $d/l_B=1.3$.
It is evident that the low-energy spectra exhibit strong anisotropy with a distinct spectra structure, and no recognizable gap separates the ground state manifold from the excited states. Notably, this phase displays a conspicuous set of quasi-degenerate states differing by a characteristic momentum along one direction, as indicated by the line connecting the lowest energy state of every momentum [see the left panel of Fig.~\ref{fig:fig1}(d)], while such a structure is absent along the other direction. These observations suggest a non-uniform real-space structure within the system, consistent with a unidirectional charge density wave phase. In the following, we further examine the charge structure factor to verify that this phase is indeed a charge stripe phase.

Due to the symmetric nature of the two layers, we compute the charge structure factor of one layer, which is defined as
\begin{equation}
\begin{aligned}
S(\mathbf{q})=\dfrac1{N_e}\langle\rho_\mathbf{q}\rho_{-\mathbf{q}}\rangle
=\dfrac1{N_e}\sum_{i,j}\langle e^{i\mathbf q\cdot{\mathbf R}_i}e^{-i\mathbf q\cdot{\mathbf R}_j}\rangle.
\end{aligned}
\end{equation}
Here, $\rho_\mathbf{q} = \sum_{i=1}^{N_e} e^{i\bold q \cdot \mathbf R_i}$ is the Fourier-transformed guiding center density and $N_e$ is the electron number of one layer.
As depicted in \figref{fig:fig3}(a), when $d<d_c$, the system appears uniform without a peak in $S(\mathbf{q})$.
However, significant peaks emerge when the layer distance exceeds the critical value $d_c$, providing clear evidence of the striped feature.  
The position of the peaks in $S(\mathbf{q})$ signifies the wave vector of such unidirectional charge order.
Within the stripe phase, the peak is precisely located at $\mathbf{q}^* = (2\cdot\frac{2\pi}{a},0)$, corresponding to a root configuration pattern (in Landau gauge) of 1111000011110000 and its translated equivalents. Remarkably, prefactor 2 in $\mathbf{q} ^*$ precisely matches the quasi-degenerate state spacing in the left panel of Fig.~\ref{fig:fig1}(d), implying the strong density-density correlations in the ground states at this charge ordering wave vector. Furthermore, the density profile (see \figref{fig:fig3}(b)) computed by VUMPS directly indicates spontaneous breaking of the translational symmetry (periodic modulation) for $d>d_c$, whereas the translational symmetry persists with uniform density in the superfluid phase at $d<d_c$. The above observations support the translational symmetry breaking at $d>d_c$.

\emph{Continuous transition at $d=d_c$. ---}
We now examine the nature of this quantum phase transition at $d=d_c$ based on distinct numerical evidence. First, the low-lying energy spectra evolve smoothly without level crossing as the layer distance increases [see \figref{fig:fig1}(a)], indicating a continuous transition. Second, the order parameter vanishes continuously across the critical point, further supporting the continuous nature of this transition. Third, both ED and VUMPS results reveal that the first-order derivative of the ground-state energy with respect to $d$ remains continuous at the critical point $d_c$  (see Supplementary Material~\cite{supplement}), conforming to the theoretical expectations for a continuous quantum phase transition. Below we present direct evidence based on ground-state fidelity.

The ground-state fidelity is defined as the wave-function overlap between the ground state at $d$ and $d+\Delta d$, i.e., $F(d,\Delta d)=|\langle\Psi(d+\Delta d)|\Psi(d)\rangle|$. It has proven to be a sensitive diagnostic for distinguishing continuous phase transitions from first-order ones, including weakly first-order transitions \cite{PhysRevE.74.031123,doi:10.1142/S0217979210056335,PhysRevE.88.032110}. In the scenario of first-order transitions, a level crossing at the transition point results in a dramatic change in the ground state and a sharp drop in fidelity~\cite{PhysRevE.88.032110}.
In contrast, a continuous quantum phase transition manifests itself as a small dip near the critical point and tends to zero as $\Delta d \to 0$.
In our infinite-system VUMPS calculations, we consider the ground-state fidelity per unit cell.
As shown in \figref{fig:fig0}(b), the VUMPS-calculated ground-state fidelity exhibits a characteristic dip near the critical layer separation $d_c \approx 0.86 l_B$. The dip weakens as the step size $\Delta d$ decreases, indicating a continuous quantum phase transition. This behavior is consistent with the ED results (see Supplementary Material~\cite{supplement}), and the VUMPS results rule out the possible finite-size artifacts. 

\emph{Conclusion and Discussions.---} 
Through systematic VUMPS and ED simulations, we establish the existence of a direct and continuous quantum phase transition in bilayer quantum Hall systems at half-filled third Landau levels (n=2). By varying the interlayer separation $d/l_B$, we observe a direct transition between two distinct symmetry-broken phases. Comprehensive evidence from energy spectrum evolution, ground-state fidelity, exciton superfluid order parameter, and derivatives of the ground-state energy unambiguously demonstrates the continuous nature of this transition. Crucially, the transition connects an exciton superfluid phase with spontaneous $U(1)$ symmetry breaking and an unidirectional charge density wave with broken translational symmetry, thereby going beyond the LGWF paradigm and establishing a concrete realization of a DQCP.

Compared with previously studied DQCP candidates involving complex spin interactions, our proposed platform offers a more straightforward experimental route with precisely tunable interactions. 
In particular, recent advances in two-dimensional materials \cite{shi_shih_rhodes_kim_barmak_watanabe_taniguchi_papić_abanin_hone_et_al._2022,https://doi.org/10.1002/advs.202300574,li_chen_wei_chen_huang_zhu_zhu_an_song_gan_et_al._2024}, such as graphene bilayers and transition metal dichalcogenides, offer promising pathways. These systems allow for precise electrostatic control of carrier density,  thereby regulating effective layer distance and realizing the tunable bilayer geometry studied here. We anticipate that the interplay between tunable interactions in such engineered 2D platforms and the universal critical behavior uncovered here will catalyze experimental verification of DQCP physics, thereby bridging a long-standing gap between quantum critical theory and solid-state implementations. Moreover, our work may also stimulate future theoretical studies, particularly the development of effective field theories that go beyond previous approaches by incorporating the topological nature of Landau levels.

\textit{Acknowledgement.---} We thank helpful discussions with Yang Liu, Yang Qi, Tongzhou Zhao, Yunlong Zang. This work was supported by the National Natural Science Foundation of China (Grant Nos. 92477106, 12488201), Innovation Program for Quantum Science and Technology (2021ZD0301800), and the Fundamental Research Funds for the Central Universities.

\bibliography{ref}%

\clearpage
\begin{appendices}
\setcounter{figure}{0}  
\setcounter{table}{0}   
\setcounter{equation}{0} 
\renewcommand{\thefigure}{S\arabic{figure}}
\renewcommand{\thetable}{S\arabic{table}}
\renewcommand{\theequation}{S\arabic{equation}}

\section{Supplementary Material for ``Deconfined Quantum Critical Point in Quantum Hall Bilayers''}

\section{Hamiltonian on a Torus}
The single-particle wave function of the nth Landau level on the torus is
\eqs{\label{eq:basis}\phi_{n,j}=\dfrac{1}{\sqrt{2^nn!\sqrt{\pi}bl_B}}\sum_{k=-\infty}^\infty H_n[\frac{1}{l_B}(X_j+ka-x)]\\
\exp[\frac{i}{l_B^2}(X_j+ka)y-\frac{1}{2l_B^2}(X_j+ka-x)^2]}
with $H_n$ the Hermite polynomials, $X_j=\frac{2\pi l_B^2}{b}j$, $l_B=\sqrt{\frac{\hbar c}{eB}}$ the magnetic length
, and $a$ and $b$ the length and width of the torus. After being projected to the n=2 Landau level, the second quantization form of the Hamiltonian of our model is
\eqs{\label{eq:2ndQ}H=\frac{1}{2ab}\sum_{\alpha}\sum_{j_1,j_2,j_3,j_4}\sum_\mathbf{q}V_{intra}(\mathbf{q})L_2^2(\frac{1}{2}q^2l_B^2)\\
\exp[-\frac{q^2l_B^2}{2}+i(X_{j_1}-X_{j_3})q_x]\\
\delta'_{j_1+j_2,j_3+j_4}\delta'_{q_y \frac{b}{2\pi},j_1-j_4}a^\dagger_{\alpha j_1}a^\dagger_{\alpha j_2}a_{\alpha j_3}a_{\alpha j_4}\\
+\frac{1}{a b}\sum_{j_1,j_2,j_3,j_4}\sum_\mathbf{q}V_{inter}(\mathbf{q})L_2^2(\frac{1}{2}q^2l_B^2)\\
\exp[-\frac{q^2l_B^2}{2}+i(X_{j_1}-X_{j_3})q_x]\\
\delta'_{j_1+j_2,j_3+j_4}\delta'_{q_y \frac{b}{2\pi},j_1-j_4}a^\dagger_{\downarrow j_1}a^\dagger_{\uparrow j_2}a_{\uparrow j_3}a_{\downarrow j_4}.}
Here, $\alpha=\uparrow,\downarrow$ is the layer index, $q=\sqrt{q_x^2+q_y^2}$, $V_{intra}(\mathbf{q})=2\pi e^2/(\epsilon q)$ and $V_{inter}(\mathbf{q})=2\pi e^2/(\epsilon q)\cdot e^{-qd/l_B}$ are the Fourier transforms of the intralayer and interlayer Coulomb interaction. $d$ denotes the layer distance
, and $L_2$ is the Laguerre polynomial for n=2. 
The $\delta'_{mn}=0$ unless $\mod(m-n,N_\phi)=0$, with the total orbital number $N_\phi=ab/(2\pi l_B^2)$.

On the torus, we exploit magnetic translation symmetry and particle number conservation. Consequently, eigenstates can be labeled by particle number $N_e$ and magnetic translation momentum $\mathbf{K}$.

\section{VUMPS on infinite cylinder}

The matrix product state (MPS) algorithms have been successfully employed to investigate quantum Hall problems in various geometries. In the sphere and finite cylinder geometries, the presence of the "shift" $\mathcal{S}$ (or equivalently, the guiding center spin)  modifies the relation between the number of flux quanta $N_\phi$ and the electron number $N_e$. For well-known states such as the Laughlin and Moore–Read states, the orbital number shift $\mathcal{S}$ is known. However, if $\mathcal{S}$ is unknown a priori, one must try different values and compare the resulting ground-state energies to identify the correct configuration for a given filling factor. 

In addition to the shift issue, geometries such as the finite cylinder and disk introduce further complications due to edge effects. Edge states in these systems potentially obstruct the attainment of the desired bulk filling factor, especially in the presence of long-range Coulomb interactions. In such cases, stabilizing the system often necessitates careful fine-tuning of the electrostatic potential at the boundary.

In contrast, the infinite cylinder geometry avoids both the shift issue and undesired edge effects, owing to its inherent translational invariance along the cylinder axis. This makes it particularly suitable for investigating intrinsic bulk properties of quantum Hall systems using MPS algorithms. 

\subsection{Uniform MPS and Symmetry Implementation}
A uniform matrix product state (uMPS) of bond dimension $\chi$ with unit cell size $q$ is defined by $q$ independent tensors $A_k^{(s)} \in \mathbb{C}^{\chi \times d \times \chi}$ ($k = 1, \dots, q$), where $s = 1, \dots, d$ labels the physical index at each site, with $d$ being the local Hilbert-space dimension. These tensors collectively define the unit cell tensor:

\begin{equation}
\mathbb{A}^{\mathbf{s}_n} = A^{(s_{nq+1})}_{nq+1} A^{(s_{nq+2})}_{nq+2} \cdots A^{(s_{nq+q})}_{nq+q},
\end{equation}

\noindent where $\mathbf{s}_n = (s_{nq+1}, s_{nq+2}, \dots, s_{nq+q})$ denotes a combined physical index for the $n$-th unit cell.
The full variational state is then written as:

\begin{equation}
|\Psi\rangle = \sum_{\{\mathbf{s}\}} \cdots \mathbb{A}^{\mathbf{s}_{n-1}} \mathbb{A}^{\mathbf{s}_{n}} \mathbb{A}^{\mathbf{s}_{n+1}} \cdots |\mathbf{s}\rangle.
\end{equation}

\subsubsection{Symmetry Considerations in Quantum Hall Bilayers}
For quantum Hall bilayer systems, we leverage three conserved quantities:
\begin{enumerate}
    \item Particle number in the upper layer
    \item Particle number in the lower layer
    \item $K_y$ momentum component
\end{enumerate}
This enables the imposition of $U(1) \otimes U(1) \otimes U(1)$ symmetry on the uMPS representation. When implementing the variational uniform matrix product state (VUMPS) algorithm, it is advantageous to define renormalized quantum numbers relative to their mean values:

\begin{align}
\tilde{C}_{i,\uparrow} &= C_{i,\uparrow} - \nu_{\uparrow} \\
\tilde{C}_{i,\downarrow} &= C_{i,\downarrow} - \nu_{\downarrow} \\
\tilde{K}_i &= K_i - i\nu,
\end{align}
where $C_{i,\uparrow/\downarrow}$ is the particle number at site $i$ in the upper/lower layer, $\nu=\nu_{\uparrow}+\nu_{\downarrow}$ is the mean particle number of the system, $\nu_{\uparrow}/\nu_{\downarrow}$ is the mean particle number of the upper/lower layer and $K_i$ is the y-component of the momentum at site $i$.

\subsubsection{Quantum Number Transformation Under Translation}
Although the tensor components $\mathbb{A}$ remain invariant under lattice translations (reflecting uniform entanglement patterns), the quantum numbers assigned to each tensor leg must transform accordingly. Under a translation by $m$ sites, the renormalized momentum quantum number undergoes the shift:

\begin{equation}
\tilde{K}_i \rightarrow \tilde{K}_i + m\tilde{C}_i,
\end{equation}
where $\tilde{C}_i=\tilde{C}_{i,\uparrow}+\tilde{C}_{i,\downarrow}$.
This transformation rule plays a crucial role in the practical implementation of the VUMPS algorithm, as it maintains consistency between local quantum numbers and global symmetry constraints during optimization.

\subsection{VUMPS Algorithm Implementation}
We implement the VUMPS algorithm beginning with an initial tensor with small bond dimensions. The left and right environments are computed directly in the thermodynamic limit by solving fixed-point equations. Unit cell tensors are then optimized through simultaneous solution of three eigenvalue problems. This iterative procedure incorporates control bond expansion (CBE)—a technique preserving the translational symmetry of uMPS to increase bond dimension until the energy and observables converge systematically.

Whereas conventional infinite DMRG (iDMRG) accumulates path-dependent memory in environmental tensors, potentially predisposing simulations to metastable states, VUMPS demonstrates superior convergence behavior despite higher computational overhead per iteration. This advantage stems from its fundamentally memory-less architecture and the concurrent solution of multiple eigenproblems.

\subsection{Cylinder Geometry Hamiltonian Formulation}
When studying quantum Hall systems in the cylinder geometry, the Hamiltonian differs slightly from that of the torus case. The single-particle basis of the nth Landau level on the cylinder is
\eqs{\label{eq:basis}\phi_{n,j}=\dfrac{1}{\sqrt{2^nn!\sqrt{\pi}bl_B}} H_n[\frac{1}{l_B}(X_j-x)]\\
\exp[\frac{i}{l_B^2}X_j y-\frac{1}{2l_B^2}(X_j-x)^2].}
Accordingly, the Hamiltonian changes to
\begin{equation}
\begin{aligned}
H=\frac{1}{2ab}\sum_{i}\sum_{j_1,j_2,j_3,j_4}\sum_{q_y}\int\frac{dq_x}{2\pi}V_{intra}(\mathbf{q})L_2^2(\frac{1}{2}q^2l_B^2)\\
\exp[-\frac{q^2l_B^2}{2}+i(X_{j_1}-X_{j_3})q_x]\\
\delta_{j_1+j_2,j_3+j_4}\delta_{q_y \frac{b}{2\pi},j_1-j_4}a^\dagger_{i j_1}a^\dagger_{i j_2}a_{i j_3}a_{i j_4}\\
+\frac{1}{a b}\sum_{j_1,j_2,j_3,j_4}\sum_{q_y}\int\frac{dq_x}{2\pi}V_{inter}(\mathbf{q})L_2^2(\frac{1}{2}q^2l_B^2)\\
\exp[-\frac{q^2l_B^2}{2}+i(X_{j_1}-X_{j_3})q_x]\\
\delta_{j_1+j_2,j_3+j_4}\delta_{q_y \frac{b}{2\pi},j_1-j_4}a^\dagger_{\downarrow j_1}a^\dagger_{\uparrow j_2}a_{\uparrow j_3}a_{\downarrow j_4}.
\end{aligned}
\end{equation}
We adopt a modified Coulomb interaction as
\begin{equation}
\begin{aligned}
    V_{\rm{intra}}(r)&= \frac{e^{-r^2/\xi^2}}{r}\\
    V_{\rm{inter}}(r)&= \frac{e^{-r^2/\xi^2}}{\sqrt{r^2+d^2}}.
\end{aligned}
\end{equation}
We retain matrix elements with amplitude $>5\times10^{-6}$, yielding MPO bond dimensions $> 1800$. One technical reason for the modification is that the Fourier transform of the interaction appears in the Hamiltonian, and the Fourier transformation of the standard Coulomb interaction has a singularity in $\mathbf{q}=0$. In torus geometry, we can just drop out the term due to the positive electric background, while in the cylinder case, the singularity is inside an integral, so we can not simply separate it. By employing a modified Coulomb interaction, all terms in the Hamiltonian have finite amplitude, so that we can consider the effect of the background after calculation, which only shifts the energy by a constant.

\begin{figure}
    \centering
    \includegraphics[width=0.9\linewidth]{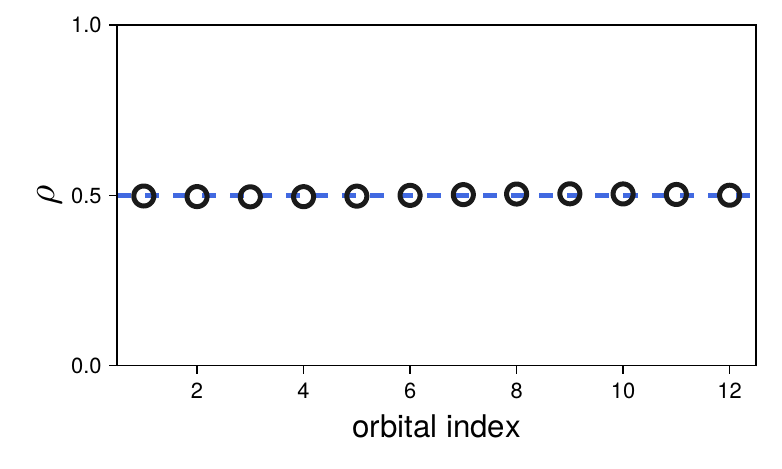}
    \caption{The Measurements of orbital occupation at $d/l_B=0.8575$ (near criticality) show constant $\rho=0.5$ across orbitals with fluctuations $\Delta \rho<10^{-2}$, verifying simulation convergence without artificial symmetry breaking.}
    \label{fig:rho}
\end{figure}

\subsection{Convergence of the VUMPS Calculation}
Near the critical point, finite bond dimension limits the ability of the matrix product state (MPS) ansatz to faithfully represent the true ground state. Specifically, it lacks the expressive power to accommodate the high entanglement entropy of the critical state. As a result, the MPS approximation may resort to artificially breaking a symmetry of the system to fit within its constrained entanglement capacity. This artificial symmetry breaking disappears as the bond dimension increases sufficiently to ensure convergence of the simulation.

To verify convergence, we measured the orbital occupation number $\rho$ at $d/l_B=0.8575$ (close to the critical point $d_c$). As shown in Fig.~\ref{fig:rho}, the converged ground state with bond dimension $\chi=4000$ exhibits uniform orbital occupation with negligible fluctuations ($\Delta \rho<10^{-2}$). This uniformity confirms that the simulation converged without artificial symmetry breaking and further excludes the possibility of an intermediate phase in which superfluid and stripe coexist.

\begin{figure}
    \centering
    \includegraphics[width=0.9\linewidth]{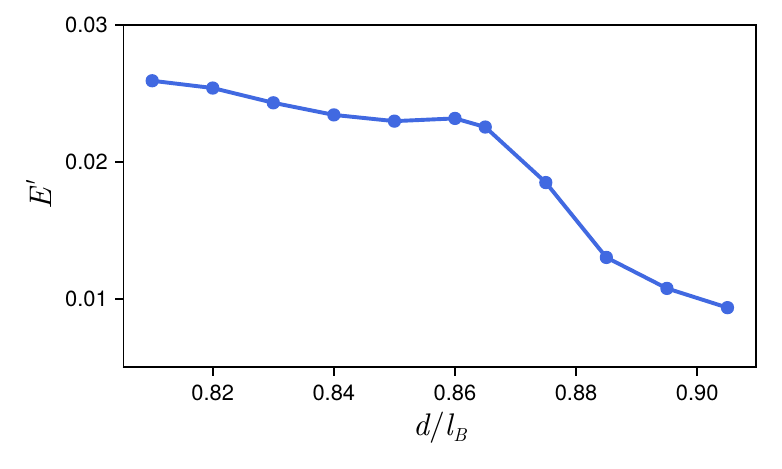}
    \caption{The first-order derivative of the ground-state energy with respect to layer distance $d/l_B$, computed via VUMPS remains continuous across the critical region.}
    \label{fig:dEdd}
\end{figure}

\section{pseudospin gap}
We consider a uniform and positive background charge to offset the divergence in the Coulomb interaction.
The divergent term in the Coulomb interaction at $\mathbf{q}=0$ is
\eqs{\label{eq:pseudospin}\frac{1}{2N_\phi}\frac{1}{q}(N_{e\uparrow}-1)N_{e\uparrow}+\frac{1}{2N_\phi}\frac{1}{q}(N_{e\downarrow}-1)N_{e\downarrow}+\\
\frac{1}{N_\phi}\frac{e^{-qd}}{q}N_{e\uparrow}N_{e\downarrow},
}
where $S_z=\frac{N_{e\uparrow}-N_{e\downarrow}}{2}$,$N_e=N_{e\uparrow}+N_{e\downarrow}$. If we simply ignore the $\mathbf{q}=0$ term, we also throw away some non-divergent constant term. Therefore, when comparing energies across different $S_z$ sectors, it is essential to include these constant terms, as they depend on $S_z$.
Substitute $S_z$ and $N_e$ for $N_{e\uparrow}$ and $N_{e\downarrow}$, \eqnref{eq:pseudospin} becomes
\begin{equation}
\begin{aligned}
&\frac{1}{2N_\phi}\frac{1}{q}(S_z+\frac{1}{2}N_e-1)(S_z+\frac{1}{2}N_e)+\\
&\frac{1}{2N_\phi}\frac{1}{q}(\frac{1}{2}N_e-S_z-1)(\frac{1}{2}N_e-S_z)+\\
&\frac{1}{N_\phi}\frac{e^{-qd}}{q}\frac{1}{4}(N_e^2-4S_z^2)\\
=&\frac{1}{4qN_\phi}[N_e^2(1+e^{-qd})-2N_e]+\frac{1}{N_\phi}\frac{1-e^{-qd}}{q}S_z^2\\
=&\frac{1}{4qN_\phi}[N_e^2(1+e^{-qd})-2N_e]+d\frac{S_z^2}{N_\phi}.
\end{aligned}
\end{equation}
The last term $d\dfrac{S_z^2}{N_\phi}$ is the constant we should add when evaluating energy differences for different $S_z$ sectors.

\begin{figure}[tbp]
\includegraphics[width=0.9\linewidth]{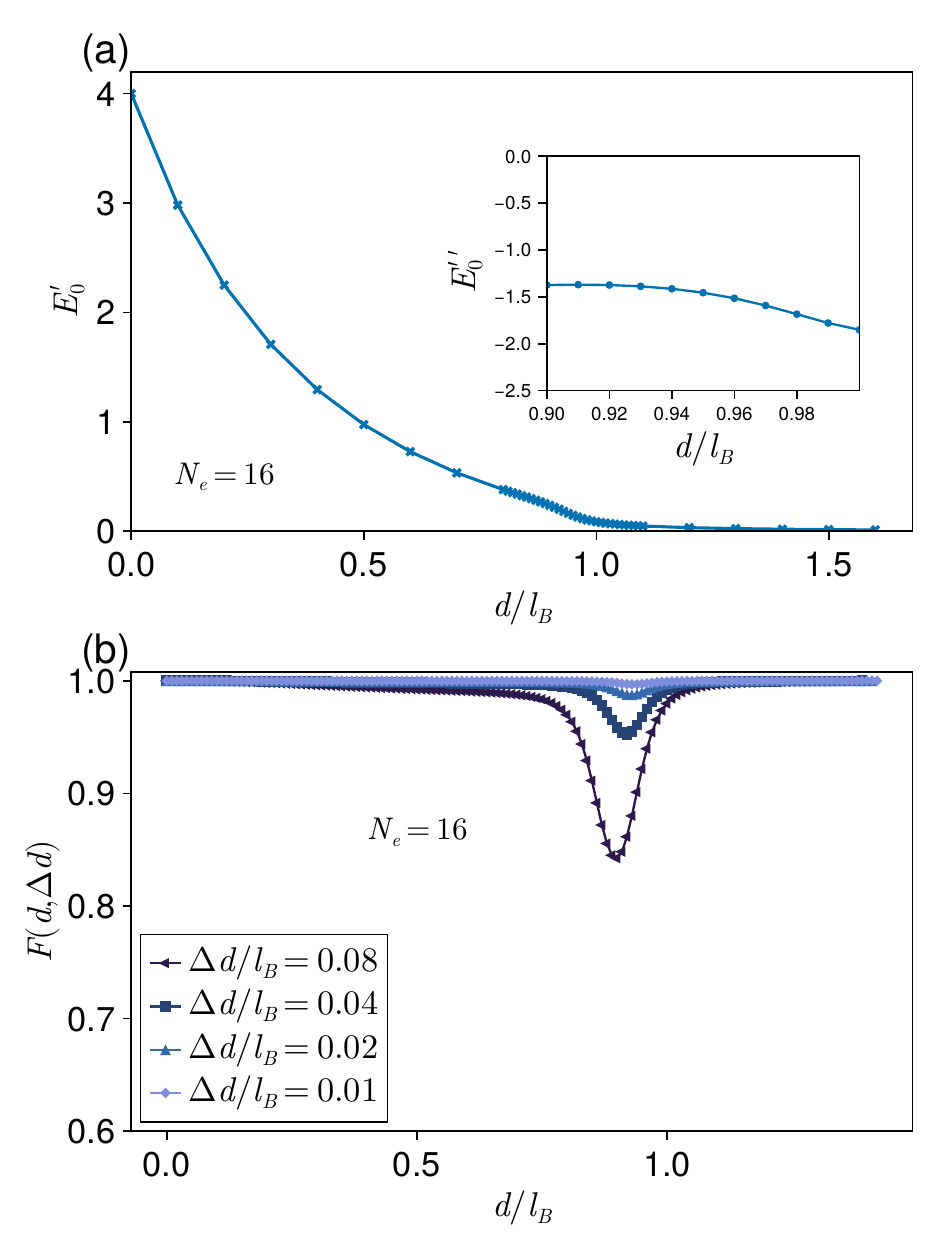}
\caption{(a) The first-order derivative of the ground-state energy with respect to layer distance $d/l_B$. The inset shows the second-order derivative. The smooth curves around $d/l_B\approx 0.93$ signal a continuous quantum phase transition. (b) The ground-state fidelity versus $d/l_B$. A small dip instead of a sudden jump appears near the critical point, suggesting a continuous quantum phase transition.}
\label{fig1ab}
\end{figure}

\section{MORE Numerical RESULTS}

\subsection{Ground-state Fidelity}
We have shown the ground-state fidelity obtained by VUMPS in the main text. We also examine it by ED, which provides consistent findings.  As shown in \figref{fig1ab}(b),  the ground-state fidelity exhibits a single weak dip at the critical distance instead of a sudden jump, and such a dip weakens with the decrease of $\Delta d$. If the transition is first-order, the fidelity will remain a finite gap from 1 once the measured points of the fidelity cross the critical point. Thus, the numerical evidence strongly indicates that the direct quantum phase transition between the two phases is continuous.

\subsection{Ground-state Energy}
In addition to the ground-state fidelity, the continuous quantum phase transition can also be verified from the derivative of the ground-state energy, and thus we examine this quantity using both VUMPS and ED.  As shown in \figref{fig:dEdd} and \figref{fig1ab}(a), the first-order derivative of the ground-state energy is continuous with layer distance $d/l_B$.
These observations support the continuity of this phase transition. 

\begin{figure}[tbp]
\includegraphics[width=0.9\linewidth]{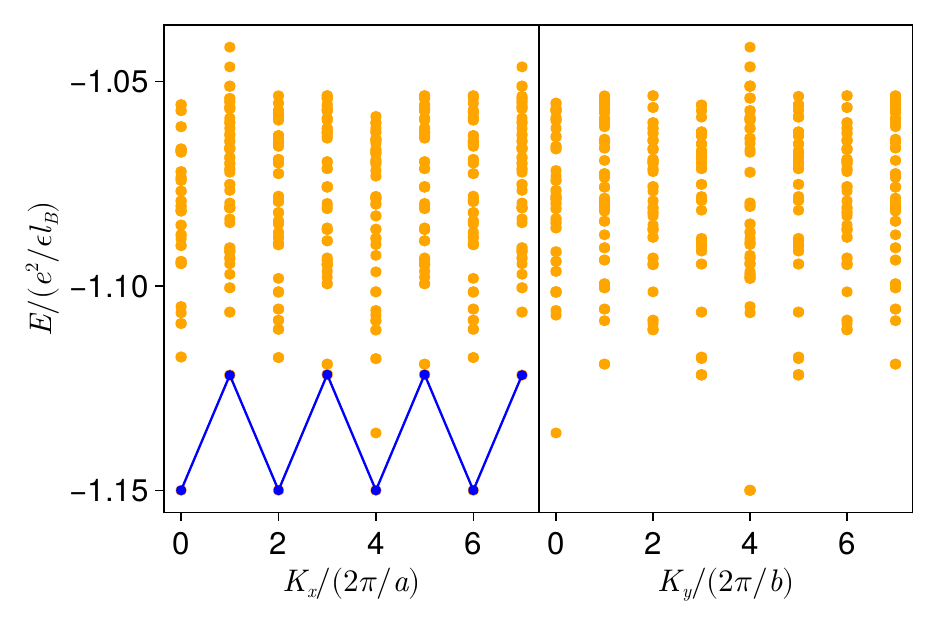}
\caption{Energy spectrum of a single layer of half-filled third Landau level. }
\label{fig1-append}
\end{figure}

\begin{figure}[tbp]
\includegraphics[width=0.9\linewidth]{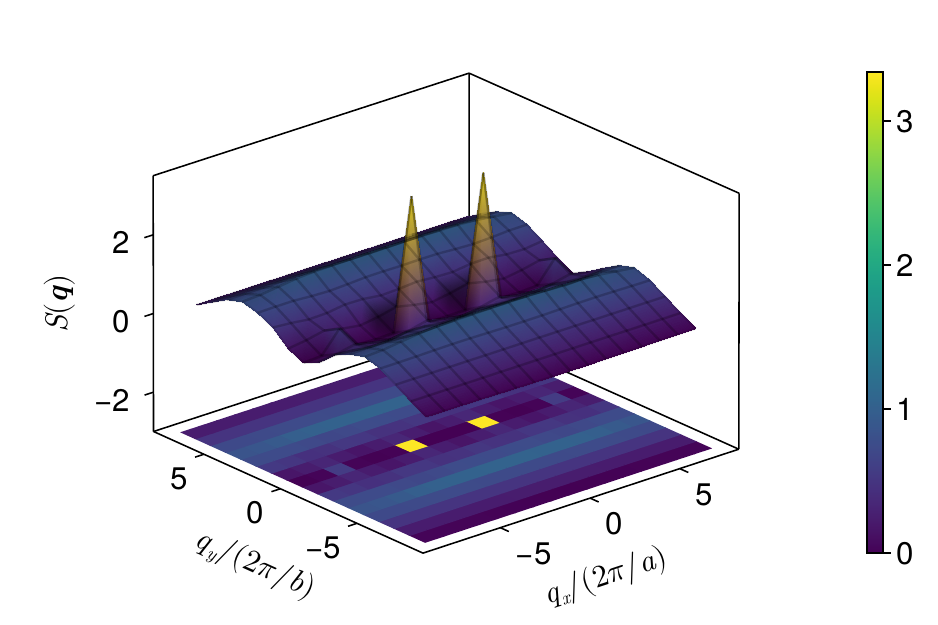}
\caption{The guiding center structure factor with $8$ electrons. The two strong peaks indicate the CDW phase in a single layer.}
\label{fig2-append}
\end{figure}

\subsection{SINGLE LAYER RESULT}
\label{append::singleLayerResult}
We calculate properties of a single layer at half-filling in the third Landau level with ED. The energy spectrum and guiding center structure factor are presented to convince the stripe phase of the decoupled bilayer.

We compute the energy spectra for the half-filled $n=2$ Landau level with eight electrons, rectangular geometry, and aspect ratio $0.64$. As shown in \figref{fig1-append}, the spectrum has the same property as the large layer distance phase in the bilayer system, which exhibits characteristic momentum difference of the groundstates in the x direction while showing no obvious feature in the y direction.

We also calculate the guiding center structure factor (as defined in the main context). In \figref{fig2-append}, the sharp peaks are consistent with the double layer result, which is located at $\mathbf{q}^*=(\pm 2\cdot\frac{2\pi}{a},0)$, suggesting a stripe phase.
\end{appendices}
\end{document}